\documentclass[aps,prd,twocolumn,floatfix]{revtex4}
\usepackage{graphicx}
\usepackage{dcolumn}
\usepackage{bm}
\begin{document}

\title{(A Few) Recent Developments in Hadron Spectroscopy}

\author{
T.Barnes$^{a,b}$\footnote{Address from 3 Jan. 2011: U.S. Department of Energy, 
Office of Nuclear Physics.}}

\affiliation{
$^a$Physics Division, Oak Ridge National Laboratory,
Oak Ridge, TN 37831-6373, USA\\
$^b$Department of Physics and Astronomy, University of Tennessee,
Knoxville, TN 37996-1200, USA}

\date{\today}

\begin{abstract}
Recent years have seen rapid developments in our knowledge and understanding
of meson spectroscopy, especially in the charm quark sectors. In my invited
overview I discussed some of these recent new developments, including
theoretical developments, new production mechanisms such as B decays and
double charmonium production, and the discovery of several of the many new
candidates for excited charmonia, charm meson molecules, and hybrid (excited glue) mesons,
in both charmonium and light quark sectors. In this writeup, due to length constraints
I will restrict my discussion to a few examples of these new states, some of their
broader theoretical implications, and future prospects.

\end{abstract}

\maketitle

\section{Introduction}

Hadron spectroscopy has undergone very rapid development in recent years, primarily as a result of
the discovery of many new states with interesting and often surprising properties, especially in
the spectroscopy of mesons containing charm quarks. In this invited overview I will discuss
a few examples of these new states, ordered by the general categories of hadrons expected by theorists.
These are 1) charmed hadrons and charmonia, 2) charm meson molecules, and 3) hybrid mesons.
As the list of new states is quite extensive ({\it ca.} 20 have been reported
and discussed), for more detailed information I refer the reader to recent reviews on the subject,
such as those by Godfrey and Olsen \cite{Godfrey:2008nc}, Swanson \cite{Swanson:2006st} and
(for exotic mesons) Meyer and van Haarlem \cite{Meyer:2010ku}.
I conclude with a summary of future prospects for the field.

\section{Charmed Hadrons and Charmonia}

The new era of spectroscopy began with the discovery of the
very narrow $J^P = 0^+$ and $1^+$ states
D$_{s0}^*(2320)$ and D$_{s1}(2460)$
\cite{Aubert:2003fg,Aubert:2003pe,Besson:2003cp,Abe:2003jk}.
Although venerable and otherwise rather accurate quark potential models predicted P-wave $c\bar s$ states
with these quantum numbers, they anticipated these states at rather higher masses, above
DK and D$^*$K thresholds, and these allowed strong decays were expected to result in
rather broad $c\bar s$ states \cite{Godfrey:1985xj}.

The initial discussions of these states included theoretical speculations that either these
were not $c\bar s$ states but were a different type of state entirely, perhaps DK and D$^*$K molecular
bound states \cite{Barnes:2003dj}, or that these were indeed $c\bar s$ states, and that
the existing potential models were simply inaccurate in their predictions \cite{Nowak:2003ra}.
Possibilities for testing the nature of these states, for example through radiative transitions,
were noted in the literature \cite{Godfrey:2003kg}.
It was also noted that proximity to the
DK and D$^*$K thresholds might have displaced initial ``bare''  $c\bar s$ valence
quark model states downwards in mass (see Hwang and Kim \cite{Hwang:2004cd} for an early
attempt to calculate this downwards mass shift). The importance of similar hadron loop effects
has long been emphasized by Tornqvist (see for example \cite{Tornqvist:1979hx}).

Although no numerically accurate model of the spectrum of $c\bar s$ states including these mass shifts
has yet been developed, this composite picture of the D$_{s0}^*(2320)$ and D$_{s1}(2460)$ as strongly mixed states with
large $c\bar s$ valence and DK (D$^*$K) continuum components now appears to be widely accepted.
Calculations of the downward mass shifts of
quark model valence basis states due to hadron loop effects have confirmed that the shifts are indeed of
the right scale. It has also been noted that even conventional charmonia experience
large downward mass shifts due to these loop effects, and that the mass shifts are similar enough to not have
been identified previously. Rather surprisingly, one may actually prove the equality of these hadron loop mass shifts
for different ``bare'' states under rather moderate assumptions (see for example \cite{Barnes:2007xu,Close:2009ii}).

In the charmonium sector, a natural first assignment of a new state would be a conventional $c\bar c$
charmonium bound state. To aid in this work, potential models of charmonia and their decays (especially their
strong decays) have now been extended to a wide range of $c\bar c$ states \cite{Barnes:2005pb,Eichten:2005ga}.
In view of the many recently discovered states that do not appear to be consistent with theoretical expectations,
it is reassuring to have one that does! This is the ``Z(3930)'', reported by Belle in two-photon collisions
\cite{Uehara:2005qd}, which is at the correct mass to be a 2P $c\bar c$ state, has $J^{PC} = 2^{++}$, and
has the $\gamma\gamma$ couplings expected for a $2{}^3$P$_2$ $c\bar c$ state; this $\chi_{c2}'(3930)$
assignment is now generally accepted.

\section{Charm Meson Molecules}

In addition to states found ``at the wrong mass'' such as the D$_{s0}^*(2320)$ and D$_{s1}(2460)$ discussed above,
there have also been recent reports of candidates for
weakly bound ``hadronic molecules''. The best known of these is the X(3872),
which was discovered by Belle in B decays to $J/\psi\pi^+\pi^-$ \cite{Choi:2003ue}.
In this case a very narrow state was reported at a mass equal to the D$^0$D$^{*0}$ threshold within experimental
errors, and the preferred quantum numbers were found to be $J^{PC} = 1^{++}$, consistent with an S-wave D$^0$D$^{*0}$
pair (antiparticle labels are suppressed). No $c\bar c$ states were anticipated with properties consistent with the X(3872)
\cite{Barnes:2003vb}, whereas well established one-pion exchange dynamics (appropriate for a weakly bound state)
indicated that this $1^{++}$ S-wave D$^0$D$^{*0}$ system might just bind \cite{Tornqvist:2003na,Swanson:2004pp}.

In this case there is a ``smoking gun'' for the molecule assignment; since a weakly-bound D$^0$D$^{*0}$ system
maximally violates isospin, decays to $J/\psi \rho^0$ (feeding the observed $J/\psi\pi^+\pi^-$ decay mode) and
$J/\psi \omega$ (feeding a $J/\psi 3\pi$ mode) with comparable branching fractions were predicted by Swanson
\cite{Swanson:2004pp}. This dramatic prediction appears to have been confirmed \cite{Abe:2005ix}, which gives
strong support to the charm meson molecule picture of the X(3872).

Perhaps the most remarkable of the new states are the ``charged charmonia'' which have been reported in
final states of a charmonium plus light hadron(s) with net charge. One example is the Z(4430), reported by the Belle
Collaboration in $\psi' \pi^{\pm}$ \cite{Choi:2007wga}. Although the status of these recently reported states is unclear,
if confirmed they would obviously be candidates for charm meson molecules, analogous to the well established X(3872).
Note that a weakly-bound two-meson molecule is different from the occasionally suggested multiquark cluster;
multiquark clusters above decay thresholds, such as the infamous pentaquark, appear unlikely because they
spontaneously dissociate or ``fall apart''.

\section{Hybrid Mesons}

Theorists have long anticipated the existence of hadrons containing both quark and gluonic excitations
\cite{Barnes:1977,Barnes:1977hg,Barnes:1982zs,Chanowitz:1982qj,Isgur:1985vy}, which are generically
referred to as ``hybrids''. Hybrid mesons have the experimentally attractive property of spanning
{\it all} $J^{PC}$ quantum numbers, including the ``exotic'' (non-$q\bar q$) $J^{PC} = 
0^{--}; 0^{+-}, 1^{-+}, 2^{+-}, ...$ that are forbidden to conventional quarkonia. Identification of such exotics
is an important part of the search for hybrids.

The lightest exotic hybrid is anticipated to have the quantum numbers $J^{PC} = 1^{-+}$, according to
both the MIT bag model \cite{Barnes:1977hg,Barnes:1982zs,Chanowitz:1982qj}
and LQCD \cite{Dudek:2009qf}.
It is interesting that the flux-tube model \cite{Isgur:1985vy} differs in anticipating {\it three} degenerate
lightest exotics, with $J^{PC} = 0^{+-}, 1^{-+}$ and $2^{+-}$.
Although two light $J^{PC} = 1^{-+}$, $I=1$ resonances were reported in LEAR and BNL (E852) data in the 1990s,
the widths of these states and complications in the analyses has led to alternative, nonresonant explanations
for these signals, with the notable exception of the robust and clearly resonant $\pi_1(1600)$ signal
in the mode $\eta' \pi$ \cite{Ivanov:2001rv}.
The COMPASS facility at CERN has recently begun to address the issue of light exotics, and
a signal for the $\pi_1(1600)$ has quite recently been reported in $\rho\pi$ through the diffractive production of
$\pi^- \pi^- \pi^+$ \cite{Alekseev:2009xt}.

A surprising development in the theory of strong decays suggests that identifying heavy-flavor hybrids
may not require the observation of exotic quantum numbers. In an LQCD study of heavy-quark hybrids,
McNeile {\it et al.} (UKQCD Collaboration) noted that decays of $Q\bar Q$-hybrids in which the heavy
$Q\bar Q$ pair remained bound, such as $H_b \to \chi_b f_0$, were dominant \cite{McNeile:2002az}.
If this applies to charmonium hybrids as well this is remarkably fortunate experimentally,
since it suggests searches for hybrid resonances in final states such as $J/\psi \pi \pi$,
rather than the much more complicated open-charm S+P final states that had previously been proposed.
This ``closed charm'' hybrid charmonium decay signature appears to be realized in the Y(4260);
this state was reported by BABAR in $e^+e^-$ annihilation into the final state $J/\psi \pi^+ \pi^-$,
through initial state radiation (ISR) \cite{Aubert:2005rm}.
Two states near 4360 and 4660 MeV with similar closed charm decay modes have since been reported in
$\psi' \pi^+ \pi^-$ final states by BABAR and Belle \cite{Aubert:2006ge,Wang:2007ea}; if confirmed
these may also be charmonium hybrid candidates.

\section{Summary and Conclusions}

We have seen a remarkable renaissance of charm hadron spectroscopy in recent years. This has first and foremost
demonstrated the importance for the field of new experimental facilities and production techniques, since
most of the new discoveries were made possible by the B factories, through the exploitation of B decays
and other newly accessible processes (e.g. double charmonium production and higher intensity two-photon collisions)
for charm hadron and charmonium production, with unprecedented statistics. Future facilities will evidently be crucial
to continue this work, and possibilities include LHCb and a super-B factory. The dedicated facilities at Beijing
(high intensity $e^+e^-$ in the charmonium region) and GSI (charmonium sector production through $p\bar p$ annihilation)
should play especially important roles in future experimental work on charm spectroscopy.

Future developments in theory are also crucial. Although potential models have proven accurate in the past, the
$c\bar s$ states have shown their limitations, and ``unquenching the quark model'' through the inclusion of hadron
loop effects is a very important next step. The accuracy and limitations of strong decay models, which are
crucial for identifying states and for estimating loop effects, is another important area for future investigation.
The existence of charm molecule candidates has demonstrated the importance of understanding hadron-hadron interactions and
identifying attractive channels which may support bound states; this topic is still poorly understood theoretically.
Perhaps the most important theoretical advance will be the application of LQCD to these areas of hadron strong decays and 
hadron-hadron interactions, since LQCD uncertainties are better understood and LQCD results can be systematically
improved.

To conclude, the previous seven years, heralded by the discovery of the unexpected narrow charm-strange mesons
D$_{s0}^*(2320)$ and D$_{s1}(2460)$, have been very exciting times indeed for hadron spectroscopy. We now have
much to understand and interpret, and can anticipate many exciting discoveries in future.

\section{Acknowledgments}

I am happy to acknowledge the kind invitation of the conference and session organizers
to present this material at ICHEP2010. I also appreciate
the opportunity to discuss the physics of hadrons with my colleagues at this meeting.
This research was supported in part by the U.S. Department of Energy under contract
DE-AC05-00OR22725 at Oak Ridge National Laboratory.
The support of the Department of Physics and Astronomy
of the University of Tennessee is also gratefully acknowledged.

\end{document}